\documentclass{egpubl}
\usepackage{STAG2025}
 
\WsPaper           % uncomment for final version of EG Workshop contribution
\usepackage[T1]{fontenc}
\usepackage{dfadobe}  

\usepackage{cite}  % comment out for biblatex with backend=biber
\BibtexOrBiblatex
\electronicVersion
\PrintedOrElectronic
\ifpdf \usepackage[pdftex]{graphicx} \pdfcompresslevel=9
\else \usepackage[dvips]{graphicx} \fi

\usepackage{egweblnk} 
\usepackage{dirtree}
\def\taskA{\emph{task A}}
\def\taskB{\emph{task B}}
\def\taskC{\emph{task C}}

\usepackage{tikz}
\usepackage{multirow}
\usepackage[normalem]{ulem}
\usepackage{xcolor}

\usepackage[caption=false]{subfig}
\usepackage[acronym]{glossaries}\glsdisablehyper
\usepackage{threeparttable}
\usepackage{makecell}
\usepackage{hyperref}
\usepackage{amssymb}
\usepackage{hhline}

\newacronym{fair}{FAIR}{Findable, Accessible, Interoperable, and Reusable}
\newacronym{guideta}{GUIDÆTA}{GUIDEd Interaction DATA}
\newacronym{gui}{GUI}{Graphical User Interface}
\newacronym{hci}{HCI}{Human Computer Interaction}
\newacronym{ir}{IR}{Information Retrieval}
\newacronym{sus}{SUS}{System Usability Scale}
\newacronym{dom}{DOM}{Document Object Model}
\newacronym{chis}{CHIS}{Consumer Health Information System}
\newacronym{apchis}{A\textsuperscript{+}CHIS}{Advanced interactive, Adaptive, personalized and visual CHIS}
\newacronym{dkt2}{DKT2}{Revised Brief Diabetes Knowledge Test}
\newacronym{cl}{CL}{Cognitive Load}
\newacronym{icl}{ICL}{Intrinsic CL}
\newacronym{ecl}{ECL}{Extraneous CL}
\newacronym{tac}{TAC}{The Attentive Cursor}

\title[GUIDÆTA]%
      {GUIDÆTA -- A Versatile Interactions Dataset\\
      with extensive Context Information and Metadata}

\author[S. Lengauer et al.]
{
\parbox{\textwidth}{%
\centering%
S. Lengauer$^{1}$\orcid{0000-0001-5136-4320}, 
S.A. von Götz$^{2}$\orcid{0009-0006-9430-8867}, 
M.T. Hoesch$^{2}$\orcid{0009-0007-3350-4570}, 
F. Steinwidder$^{1}$\orcid{0009-0004-5337-1195}, 
M. Tytarenko$^{1}$\orcid{0009-0001-6925-272X}, 
M.A. Bedek$^{2}$\orcid{0009-0005-1397-5186} and 
T. Schreck$^{1}$\orcid{0000-0003-0778-8665}}
        \\
{\parbox{\textwidth}{\centering% 
$^1$Graz University of Technology, Institute of Visual Computing, Austria\\
$^2$University of Graz, Department of Psychology, Austria
       }
}
}
\begin{document}
\setlength{\fboxsep}{0pt}

\maketitle
%-------------------------------------------------------------------------
\begin{abstract} 
Interaction data is widely used in multiple domains such as cognitive science, visualization, human computer interaction, and cybersecurity, among others. 
Applications range from cognitive analyses over user/behavior modeling, adaptation, recommendations, to (user/bot) identification/verification. 
That is, research on these applications -- in particular those relying on learned models -- require copious amounts of structured data for both training and evaluation. 
Different application domains thereby impose different requirements. 
I.e., for some purposes it is vital that the data is based on a guided interaction process, meaning that monitored subjects pursued a given task, while other purposes require additional context information, such as widget interactions or metadata. 
Unfortunately, the amount of publicly available datasets is small and their respective applicability for specific purposes limited. 
We present \gls{guideta} -- a new dataset, collected from a large-scale guided user study with more than 250 users, each working on three pre-defined information retrieval tasks using a custom-built consumer information system.
Besides being larger than most comparable datasets -- with 716 completed tasks, 2.39 million mouse and keyboard events (2.35 million and 40 thousand, respectively) and a total observation period of almost 50 hours -- its interactions exhibit encompassing context information in the form of widget information, triggered (system) events and associated displayed content. 
Combined with extensive metadata such as sociodemographic user data and answers to explicit feedback questionnaires (regarding perceived usability, experienced cognitive load, pre-knowledge on the information system's topic), \gls{guideta} constitutes a versatile dataset, applicable for various research domains and purposes. 
Alongside the data itself, we publish the software tools we use for handling and analyzing the dataset.

%-------------------------------------------------------------------------
%  ACM CCS 1998
%  (see https://www.acm.org/publications/computing-classification-system/1998)
% \begin{classification} % according to https://www.acm.org/publications/computing-classification-system/1998
% \CCScat{Computer Graphics}{I.3.3}{Picture/Image Generation}{Line and curve generation}
% \end{classification}
%-------------------------------------------------------------------------
%  ACM CCS 2012
   % (see https://www.acm.org/publications/class-2012)
%The tool at \url{http://dl.acm.org/ccs.cfm} can be used to generate
% CCS codes.
%Example:


\end{abstract}  

%-------------------------------------------------------------------------
\section{Introduction}
Interaction data in the form of mouse and keyboard events or widget interaction (the target of such an event)  contain vital information about users interacting with \glspl{gui}.
Its core advantage is that it can be collected unobtrusively and inexpensively without interfering with users' natural behavior. 
As such, it constitutes the basis for different objectives of different research domains. 
At the same time, it plays a vital role in many web-based systems where collected interaction data is used for fingerprinting, profiling, user classification/-authentication, inferring interest and other purposes.

In the field of cognitive science, behavioral indicators such as entropy, velocity, acceleration, or curvature are computed to infer users' mental activity, visual attention, cognition, emotion (e.g., stress), or interest~\cite{meyer_mouse_2023, 10.3389/fnhum.2020.565664}. 
Closely related are the objectives of the \gls{hci} domain, where similar metrics based on interaction speed and timing are leveraged to improve \glspl{gui}~\cite{meyerSpeedAccuracyTradeoffs1990,khanMouseDynamicsBehavioral2024}. 
Within the subfield of \emph{interaction ergonomics}, researchers attempt to answer questions such as ``how quickly can a user position the mouse and click on a button or a drop-down menu?'', relating to Hick's~\cite{hickRateGainInformation1952} and Fitts's~\cite{fittsInformationCapacityHuman1954} laws of experimental psychology. 
Behavioral patterns are also used to determine self-efficacy, risk-perception, willingness to learn, or perceived usefulness/ease of use~\cite{katerinaMouseBehavioralPatterns2018}.
On a similar note, in visualization (user-adaptive visualization in particular), interaction data constitutes the core input for user modeling, which -- in turn -- is used to adapt \glspl{gui} according to respective users' needs~\cite{yanez_state_2025}. 
Such adaptions can manifest
in a change of complexity, representation, or displayed content, among others~\cite{gotz_behavior-driven_2009,shaoVisualDocumentExploration2025}.

In the domain of cybersecurity, interactions are used to uniquely identify users based on their behavior. 
This subfield known as \emph{behavioral biometrics} is studied for the purpose of user authentication~\cite{khan_authenticating_2021,antal_user_2019,shen_continuous_2012} and fending against bots~\cite{jcp3010007,acien_becaptcha-mouse_2022,sun_deep_2019}. 
In essence, respective approaches build upon behavioral characteristics, such as mouse dynamics (the mouse behavior on \glspl{gui}), keystroke dynamics, swipe dynamics, widget interaction among others.

The common thread of these research domains is the need for large (real-world) datasets, containing all the necessary properties. 
With our \gls{guideta} dataset we aim to address this need. 
Collected through a purely unsupervised online study, it exhibits a broad spectrum of user demographics in terms of age and education level.
That is, its versatility enables its use for different research objectives and applications. 
In summary, the core strength of the \gls{guideta} dataset is fourfold:
\begin{enumerate}
    \item It contains exclusively \emph{guided} interaction data, meaning that study participants completed the same set of \gls{ir} tasks for which we also report whether they have been completed successfully and correctly. 
    \item As opposed to most related datasets, it also includes the context of interaction events. 
    Besides the widget (the target of an interaction), we define a set of 18 uniquely identifiable actions, which are supported by the system used for collecting the dataset. 
    A particular event can thus have the attached information that it, e.g., triggered the pop-up of a thumbnail with specific content.
    \item We provide various additional metadata. 
    On a participant level, this includes sociodemographic information and answers to different explicit feedback questionnaires, such as the system's perceived usability and pre-knowledge on the system's information domain.  
    On a per-completed-task level, we have participants' feedback on the experienced \gls{cl} and the required working time.
    \item All of the collected data is made publicly available through the \href{https://creativecommons.org/licenses/by/4.0/}{Creative Commons
Attribution (CC BY)} license, respecting the \gls{fair} Guiding Principles for scientific data management and stewardship. 
    In line with this mantra, we also provide all the accompanying tools for loading, filtering and analyzing the data. 
\end{enumerate}
In the following, we present a delineation to other related datasets (Sec.~\ref{sec:related-work}), before the collection process and the dataset's structure is described in depth (Sec.~\ref{sec:clt-dataset}). 
Sec.~\ref{sec:data-handling} explains our purposely-developed data handling and analysis tools, while Sec.~\ref{sec:discussion} gives an outlook on potential limitations, applications and follow-ups.

%-------------------------------------------------------------------------
\section{Related Work}\label{sec:related-work}

Interaction data has been studied since the early days of \glspl{gui}, pursuing different objectives. 
While the analysis of the keystroke dynamics constitutes a research focus since the 1970s, mouse dynamics experienced increased attention since the beginning of the 21st century due to the immense growth of the Internet~\cite{khanMouseDynamicsBehavioral2024}. 
Regardless of the research domain (cognitive sciences, \gls{hci}, adaptive visualization, or cybersecurity), this interest cultivates a need for (large-scale) datasets, which allow researchers to objectively evaluate hypotheses and validate models. 
Depending on the intended use, the collection process for compiling said datasets varies. 
I.e., interactions can be collected in a purely unsupervised manner over the Internet, allowing for a huge outreach while keeping costs low, or they can be collected in laboratory setups, allowing to tap into additional information channels such as eye tracking or verbal interviews. 
Orthogonal to that, participants can be asked to pursue specific tasks (guided) or just engage in their day-to-day activities (unguided)~\cite{antal_user_2019}. 
While the prior is crucial to study between-subjects' activity, cognition, or stress~\cite{meyer_mouse_2023}, the latter is completely sufficient for user authentication/identification~\cite{khan_authenticating_2021,antal_user_2019,shen_continuous_2012}. 
Kahn et al.~\cite{khanMouseDynamicsBehavioral2024} go even one step further and define 4 tiers of guidance: (1) fixed static sequence of actions, (2) app restricted continuous data collection, (3) app agnostic semi-controlled data collection, and (4) completely free data collection.

As the focus of this paper is not on specific application domains, but datasets, we discuss the most relevant publicly available datasets in the following. 
Other datasets were compiled to evaluate approaches for user modeling/behavior prediction~\cite{guoPredictingWebSearcher2010a,huangUserSeeUser2012,chenUserSatisfactionPrediction2017,maoEstimatingCredibilityUser2014} or for user authentication~\cite{hinbarjiDynamicUserAuthentication2015,zhengEfficientUserVerification2016,feherUserIdentityVerification2012} but are, unfortunately, not publicly available. 
\begin{description}
    \item [Balabit]
        Published in 2016 by F\"{u}l\"{o}p et al.~\cite{fulop_balabit_nodate}, the Balabit dataset contains mouse pointer positioning and timing information of 10 users who connected over a remove server. 
        The authors reason that participants can be uniquely identified, purely based on these mouse dynamics, preventing unauthorized usage of their accounts. 
        Consequently, the collected data is unguided in nature, as users were asked to perform their regular daily duties.
    \item [Chao Shen]
        Similarly, Shen et al.~\cite{shen_continuous_2012} asked their participants to pursue their daily work while they tracked mouse movements in the background. 
        That is, they were able to collect data pertaining to 28 individuals over a period of over two months. 
        Each of those completed at least 30 separate sessions of about thirty minutes. 
    \item [Bogazici] 
        In 2021 Kılıç et al.~\cite{kilic_bogazici_2021} published a dataset with unguided mouse interactions of 24 users with several days of active usage. 
        Besides positions and timestamps, they also collected window name (i.e., widget interaction) and mouse action details. 
    \item [DFL] 
        Antal and Denes-Fazakas~\cite{antal_user_2019} compiled a dataset of 21 different users, which they use for user verification. 
        They reason that about 60 minutes of interaction data is necessary to model a user, while at least 10 mouse actions are required for identity prediction. 
        Data collection was enabled through a background service, which users were asked to install before pursuing their daily work routine.
    \item [SapiMouse]
        The SapiMouse Dataset%
        % \footnote{\url{https://github.com/margitantal68/sapimouse}}
        , which was collected at Sapientia University in 2020 by Antal et al.~\cite{antal_sapimouse_2021,antal_sapiagent_2021} pursues a different objective. 
        As opposed to long observations over the course of days or months, they have a short fixed-length observation period of 4 minutes per user.
        However, within this time frame participants were asked to perform very concrete actions in the form of a mini game. 
        The interactions, collected from 120 participants, are thus highly guided, allowing for in-depth between-subject comparisons. 
        Based on that, they present a deep learning approach for user authentication, which learns descriptive features directly from the raw data~\cite{antal_sapimouse_2021}. 
        As a follow-up, they also present an autoencoder-based approach for generating human-like trajectories~\cite{antal_sapiagent_2021}.
    \item [\gls{tac}]
        The largest dataset to date in terms of participants, was published in 2020 by Leiva and Arapakis~\cite{10.3389/fnhum.2020.565664}.  
        They collected mouse dynamics of 2,909 subjects performing a transactional search task, together with attention labels and demographic attributes. 
        Alongside timestamp, position, and event type, they also report widget information in the form of the DOM element related to an event.    
    \item [ReMouse]
        Sadeghpour and Vlajic~\cite{jcp3010007} compiled a dataset for the very specific purpose of fending against session-replay bots. 
        Given this particular objective, the unique characteristic of their dataset is that it contains repeated sessions generated by the same human user. 
        Over the course of two days, they collected interaction data from 100 subjects participating in a `Catch Me If You Can!' online game, who were recruited over Amazon's MTurk platform. 
    \item [AdSERP]
        Most recently, Latifzadeh et al.~\cite{latifzadeh2025versatile} presented the AdSERP dataset, which combines mouse movement data with eye tracking.
        They aim to study user attention and purchasing behavior on search engine result pages. 
        To this end, they tracked the interactions of 47 participants in a supersized setting with interaction sessions of up to one minute. 
\end{description}

Table~\ref{tab:datasets-specifics} shows a detailed breakdown of the specifics of the above-mentioned datasets, including 
the total number of users, 
the total number of recorded interactions
the total observation period, 
whether they incorporate widget information, 
and the level of guidance during the collection process.
We can observe, that solely \textbf{\gls{tac}} boasts more participants, but on the downside this dataset comprises only very short interactions -- having thus a substantially shorter observation period and substantially fewer overall interaction events. 
Regarding the sheer observation duration, there are three datasets (\textbf{Balabit}, \textbf{Bogazici}, \textbf{DFL}) standing out with durations of multiple weeks or even months. 
However, these interactions stem from collection efforts which were conducted unguided and unobtrusively, asking participants to conduct their everyday work while a tracking service collected interactions in the background. 
With respect to these alternatives, 
our \gls{guideta} dataset fills a n\"iche as it exhibits a high count of participants, considering that it was collected from a user study with a high-level of guidance through purposely posed \gls{ir} tasks. 
A mean observation period of 11:46 minutes ($\mathit{SD}=$9:57 minutes) per user, allowed us to incentivize a comparably large number of participants to partake in our study while still having a reasonable amount of data per person.

\bgroup
\def\dotSize{.8ex}
\definecolor{tierdcolor}{rgb}{0.96,0.94,0.97}
\definecolor{tierccolor}{rgb}{0.74,0.79,0.88}
\definecolor{tierbcolor}{rgb}{0.40,0.66,0.81}
\definecolor{tieracolor}{rgb}{0.01,0.51,0.54}
\def \tierD {\tikz\draw[black,fill=tierdcolor] (0,0) circle (\dotSize);}
\def \tierC {\tikz\draw[black,fill=tierccolor] (0,0) circle (\dotSize);}
\def \tierB {\tikz\draw[black,fill=tierbcolor] (0,0) circle (\dotSize);}
\def \tierA {\tikz\draw[black,fill=tieracolor] (0,0) circle (\dotSize);}
\begin{table}[ht!]
    \centering
    \caption{A breakdown of related datasets in terms of scale.}
    \label{tab:datasets-specifics}
    \begin{threeparttable}
        \begin{tabular}{l r r@{.}l r@{.}l c c}
            \hline 
            Dataset 
                & \rotatebox{70}{\# Users} 
                & \multicolumn{2}{c}{\rotatebox{70}{\# Interactions}}
                & \multicolumn{2}{c}{\rotatebox{70}{\makecell{Observation\\duration\tnote{\dag}}}}
                & \rotatebox{70}{\makecell{Widget\\Information}}
                & \rotatebox{70}{Guided\tnote{\ddag}} \\
            \hline \hline
            \textbf{Balabit} & 10 & 4 & 61 M & 178 & 21 H & & \tierD \\
            \textbf{Chao Shen} & 28 & 79 & 69 M & 499 & 00 H & & \tierD \\ 
% Total number of data rows: 72827105
% Total duration: 6916251.569000244 sec
% Adjusted duration: 1942648.9360120296 sec
            \textbf{Bogazici} & 24 & 72 & 83 M & 539 & 62 H & \checkmark & \tierD \\ 
            \textbf{DFL} & 21 & 129 & 56 M & 718 & 06 H & & \tierD \\
            \textbf{SapiMouse} & 120 & 1 & 18 M & 8 & 22 H & N.A. & \tierA \\
            \textbf{\gls{tac}} & 2,909 & 0 & 11 M & 12 & 74 H & \checkmark & \tierC \\
            \textbf{ReMouse} & 100 & 0 & 38 M & 6 & 74 H & N.A. & \tierA \\ 
            \textbf{AdSERP} & 47 & 1 & 17 M & 16 & 39 H & & \tierC \\
            \textbf{\gls{guideta}} & 253 & 2 & 39 M & 49 & 61 H & \checkmark & \tierB \\
            \hline
        \end{tabular}
        \begin{tablenotes}
            \item[\dag] The accumulated duration over all sessions. As there is no uniform definition of `session', we assume that a gap of more than 10 seconds in the recording indicates a session separator. Note, that this can result in durations that substantially differ from the authors' reports. 
            \item[\ddag] Level of guidance as defined by Kahn et al.~\cite{khanMouseDynamicsBehavioral2024}: \tierA~\emph{Fixed static sequence of actions}, \tierB~\emph{App restricted continuous}, \tierC~\emph{App agnostic semi-controlled}, and \tierD~\emph{Completely free}.
        \end{tablenotes}
    \end{threeparttable}
\end{table}
\egroup

%-------------------------------------------------------------------------
\section{The \gls{guideta} Dataset}\label{sec:clt-dataset}

In the following, we provide details on the data collection process (Sec.~\ref{sec:data-collection}), used to compile the \gls{guideta} dataset. 
Sec.~\ref{sec:statistics} provides a statistical overview of the data in terms of different users and tasks, before we conclude the section with a description of the data structure we devised to store and distribute the data (Sec.~\ref{sec:data-structure}). 

%-------------------------------------------------------------------------
\subsection{Data Collection}\label{sec:data-collection}

\begin{figure}[ht!]
    \centering
    \includegraphics[width=\linewidth]{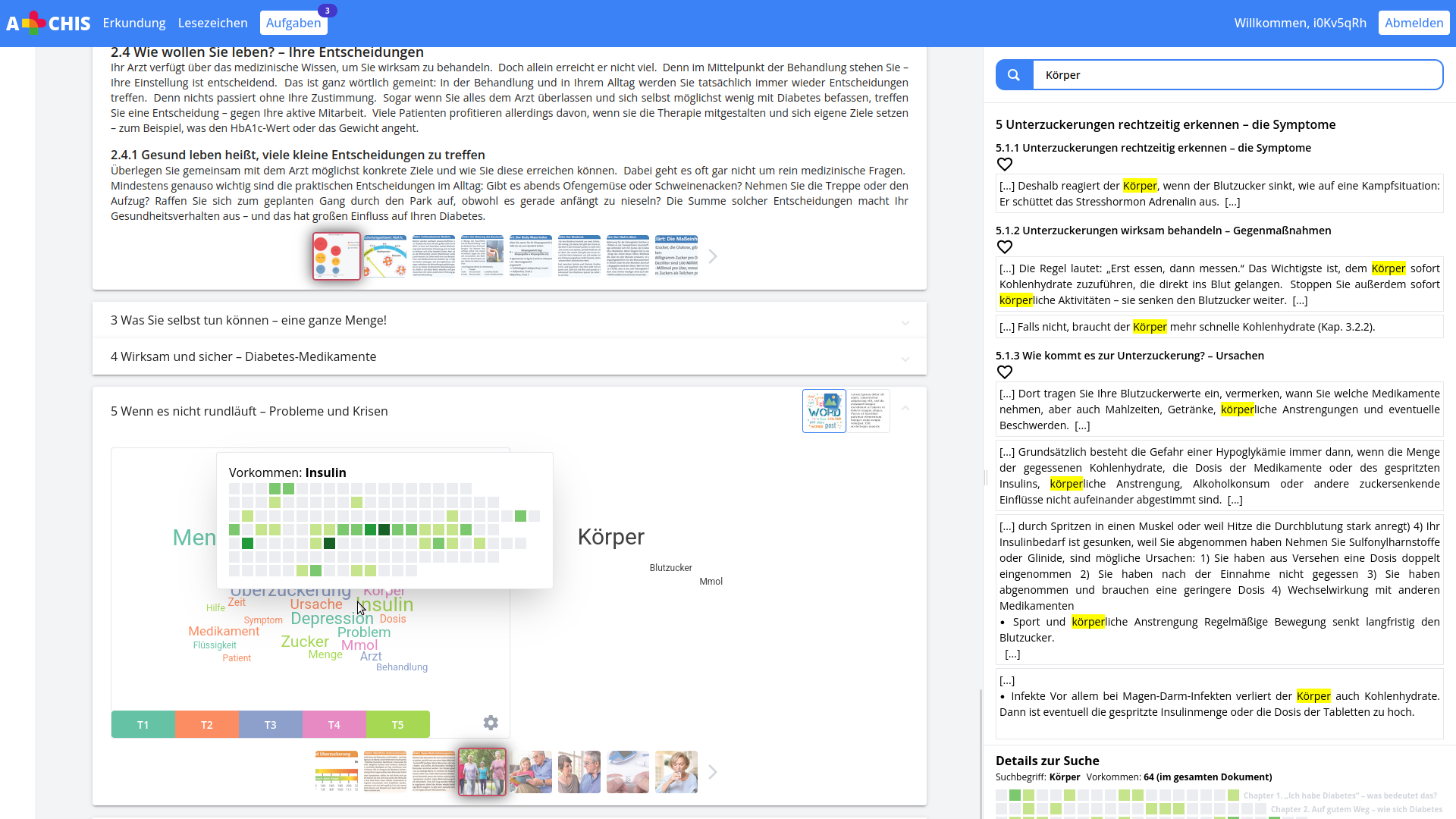}
    \caption{Screenshot of the \gls{apchis}, used in the data collection.}
    \label{fig:apchis-exploration-page}
\end{figure}

The interaction data was collected in the course of a cognitive science study from May 12 till June 23, 2025. 
\paragraph*{Study Design.}
The basis for the collection setup is a \gls{chis}, which is used by study participants to process the posed information retrieval tasks. 
This \gls{apchis}~\cite{shaoVisualDocumentExploration2025}, developed within the eponymous funding project, comprises various components for presenting the textual and pictorial content of information sources for medical knowledge (Fig.~\ref{fig:apchis-exploration-page}). 
Text can be consumed in original format but also through various \emph{distant reading}~\cite{morettiGraphsMapsTrees2005} solutions, such as word clouds, topic modelings, heat maps, search functionality, etc., which allow to process the given information in non-linear fashion. 
Subsystems are in place to display pictorial content, including interactive infographics. 

To increase the outreach, the study was conducted in purely unsupervised fashion, by providing participants with a link to the website on which they conducted the tasks using their own hardware devices. 
To foster interest in participation, advertisements were placed on the project's website, posted on social media, emailed to students and those expressing interest in the project, and distributed in physical form as posters. 
Actual ambition to solve the posed \gls{ir} tasks was incentivized with an unconditional expense allowance of €20 and the chance to get an additional €100 (which was distributed ten times by lot to all participants who were able to correctly answer all tasks).
To ensure a smooth execution they had to go through a number of well-defined steps:
\begin{enumerate}
    \item Initially, participants are greeted with a welcome page outlining the objectives of the study, clarifying the scope and what data will be collected. 
    Only after agreeing to these terms and conditions they are forwarded to the next step.
    \item Participants are presented with a form for filling out anonymous sociodemographic information and asked to complete a standardized knowledge-test on diabetes -- the \gls{dkt2}~\cite{fitzgerald2016validation} -- (the thematic domain of the \gls{apchis}).
    \item Next, they are presented with a click-through help wizard, educating them on the different components of the system through textual descriptions combined with visual examples. 
    \item Next, a task modal pops up, displaying the first of 3 tasks which are provided in randomized order: 
    \begin{description}
        \item [\taskA:] \textit{``What is the time frame after administration when the peak effect of regular insulin is reached?''},
        \item [\taskB:] \textit{``From what blood sugar level are acute measures necessary to prevent a risky hyperglycemia?''}, and
        \item [\taskC:] \textit{``How often should the HbA1c value be checked by a doctor at a minimum?''}
    \end{description}
    A participant can re-open this modal and the help wizard at any moment through distinct buttons. 
    For completing a task, the modal has to be reopened and the answer filled in. 
    After each task, participants are prompted with a form evaluating the experienced \gls{cl}, with a \gls{cl} questionnaire~\cite{krieglstein2023development}. 
    \item After completing all the tasks, a form for a \gls{sus}\cite{brookeSUSQuickDirty1996} and further details on compensation are presented to participants.
\end{enumerate}

After the study part, participants were still able to use the system for further exploration. 
We deliberately decided to pose open questions, which had to be answered in a free-form text, to prevent users from guessing. 
\begin{figure}[htb!]
    \centering
    \includegraphics[width=\linewidth]{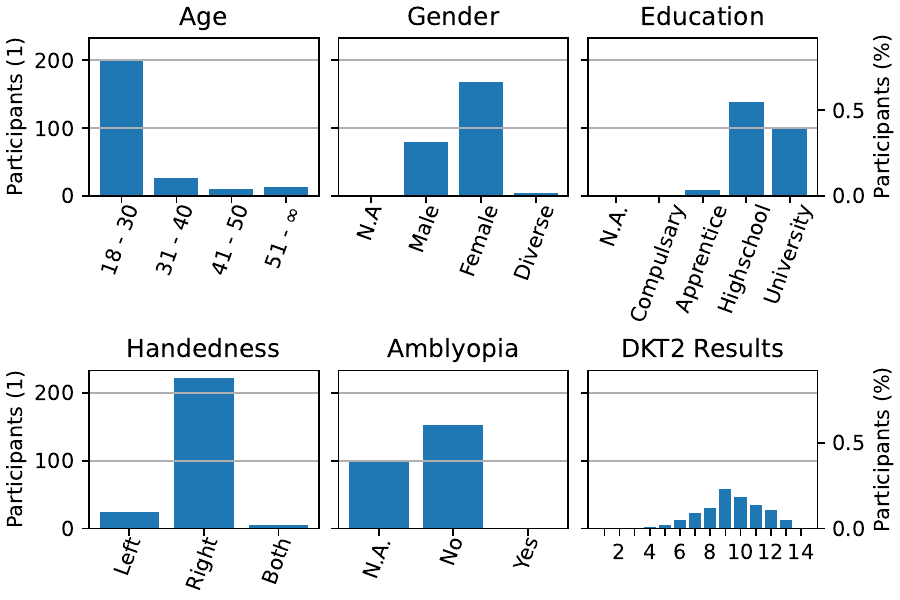}
    \caption{The distribution of age, gender, education level, handedness, amblyopia, \gls{dkt2} results among the study participants. 
    }
    \label{fig:sociodemographic-statistics}
\end{figure}

\paragraph*{Technological Setup.}
For the study the \gls{apchis} was striped down to contain only its relevant core parts. 
Yet, additional subsystems for the onboarding, task handout, questionnaires and help pages were developed to allow for a fully unsupervised execution. 
The most substantial adaptations pertain to the changes related to the comprehensive tracking of interactions. 
To this end, we employ various JavaScript event listeners\footnote{\url{https://developer.mozilla.org/en-US/docs/Web/API/Event}}, monitoring mouse events (click, scroll, move), keyboard events and window events (focus, blur, resize). 
To capture additional context, we enrich the \gls{apchis}'s \gls{dom} with custom tags labeling tools and content. 
This allows us, e.g., to reconstruct that a mouse click happened using a specific tool while exploring a specific information unit (e.g., a sentence, an image, etc.). 
All interactions are cached in the frontend and sent in batches~\cite{jcp3010007} (every 10 seconds or when a session is finalized) to a django\footnote{\url{https://www.djangoproject.com/}} backend, where they are stored in a PostgreSQL\footnote{\url{https://www.postgresql.org/}} database. 

\bgroup
\def\colsep{1mm}
\begin{table}[ht!]
    \centering
    \caption{Task related statistics such as duration or number of mouse moves. Durations are given as fractions of minutes. Relative values (e.g., mouse moves per second) are in parentheses. }
    \label{tab:statistics}
    \begin{threeparttable}
    \small{
    \begin{tabular}{@{}l@{\hskip \colsep}|@{\hskip \colsep}r@{ (}r@{)\hskip \colsep}r@{ (}r@{) }r@{ (}r@{) }r@{ (}r@{) }r@{ (}r@{) }}
    
        \hline
        Prop. & \multicolumn{2}{c}{Min} & \multicolumn{2}{c}{Max} & \multicolumn{2}{c}{Mean} & \multicolumn{2}{c}{StdDev} & \multicolumn{2}{c}{Median} \\
        \hline 
\hline
\multicolumn{10}{l}{\taskA} \\ 
\hline
Dur. & 0.31 & --- & 95.83 & --- & 6.58 & --- & 8.70 & --- & 4.30 & --- \\
Moves & 10 & 0.09 & 19.37K & 95.74 & 3.46K & 11.83 & 3.29K & 9.84 & 2.28K & 9.93 \\
Clicks & 2 & 0.01 & 238 & 0.74 & 31 & 0.12 & 31 & 0.11 & 20 & 0.09 \\
Scrolls & 0 & 0.00 & 13.13K & 18.77 & 1.03K & 2.73 & 2.02K & 3.77 & 254 & 0.81 \\
Keys & 0 & 0.00 & 1.64K & 23.71 & 72 & 0.30 & 153 & 1.55 & 38 & 0.12 \vspace{2mm} \\
\multicolumn{10}{l}{\taskB} \\ 
\hline
Dur. & 0.26 & --- & 27.78 & --- & 3.21 & --- & 3.30 & --- & 2.25 & --- \\
Moves & 8 & 0.11 & 17.43K & 86.46 & 2.35K & 13.80 & 2.56K & 9.18 & 1.60K & 11.61 \\
Clicks & 0 & 0.00 & 229 & 0.76 & 21 & 0.13 & 27 & 0.12 & 13 & 0.09 \\
Scrolls & 0 & 0.00 & 4.47K & 14.45 & 454 & 2.30 & 795 & 3.05 & 111 & 0.72 \\
Keys & 0 & 0.00 & 6.41K & 28.30 & 48 & 0.26 & 414 & 1.83 & 11 & 0.07 \vspace{2mm} \\
\multicolumn{10}{l}{\taskC} \\ 
\hline
Dur. & 0.19 & --- & 12.68 & --- & 2.67 & --- & 2.05 & --- & 2.16 & --- \\
Moves & 8 & 0.21 & 14.86K & 57.99 & 2.15K & 14.09 & 2.19K & 8.93 & 1.58K & 12.25 \\
Clicks & 1 & 0.01 & 167 & 0.66 & 22 & 0.13 & 31 & 0.11 & 12 & 0.10 \\
Scrolls & 0 & 0.00 & 3.64K & 15.76 & 350 & 2.13 & 562 & 2.94 & 114 & 0.86 \\
Keys & 0 & 0.00 & 7.01K & 28.94 & 44 & 0.25 & 453 & 1.87 & 7 & 0.07 \vspace{2mm} \\
\multicolumn{10}{l}{$\sum$\tnote{\dag}} \\ 
\hline
Dur. & 2.16 & --- & 71.34 & --- & 11.74 & --- & 7.79 & --- & 9.90 & --- \\
Moves & 94 & 0.14 & 44.98K & 81.00 & 7.76K & 12.31 & 5.92K & 8.64 & 6.33K & 10.63 \\
Clicks & 9 & 0.02 & 423 & 0.59 & 74 & 0.13 & 57 & 0.11 & 58 & 0.09 \\
Scrolls & 0 & 0.00 & 16.10K & 15.62 & 1.88K & 2.74 & 2.78K & 3.38 & 512 & 0.91 \\
Keys & 0 & 0.00 & 14.84K & 28.07 & 171 & 0.28 & 986 & 1.86 & 74 & 0.11 \\
\hline
    \end{tabular}}
    \begin{tablenotes}
        \item[\dag] This includes only participants who completed all tasks.
    \end{tablenotes}
    \end{threeparttable}
\end{table}
\egroup

\paragraph*{Data Filtering.}
The database in its raw form is not fit for direct interaction analysis as it contains mostly system-related overhead (content, logs, temporary data, settings, etc.), but all relevant interaction data was carefully extracted and transformed into an interoperable and reusable~\cite{wilkinson2016fair} format (Sec.~\ref{sec:data-structure}). 
This was done mostly automatically, but the correctness of the users' given answers to the \gls{ir} tasks was evaluated manually. 
In this migration process, we also filtered out incomplete, irrelevant, or erroneous data. 
That is, we removed all exploration sessions with a duration of one second or less, which likely resulted from window tabbing or similar interface actions. 
This was the case for 2\% of sessions, affecting about 2\textperthousand\ of interaction events. 
We similarly treated all sessions without a finalization context -- i.e., cases where the connection was interrupted, the browser window closed unexpectedly, etc. 
Lastly, we discarded all user tasks with a duration of less than 10 seconds, as we reason that the task was not properly attempted in such cases. 
All data, orphaned by these filters (e.g., users without any tasks or mouse events without a session) are discarded as well.

\begin{figure}[ht!]
    \centering
    \includegraphics[width=\linewidth]{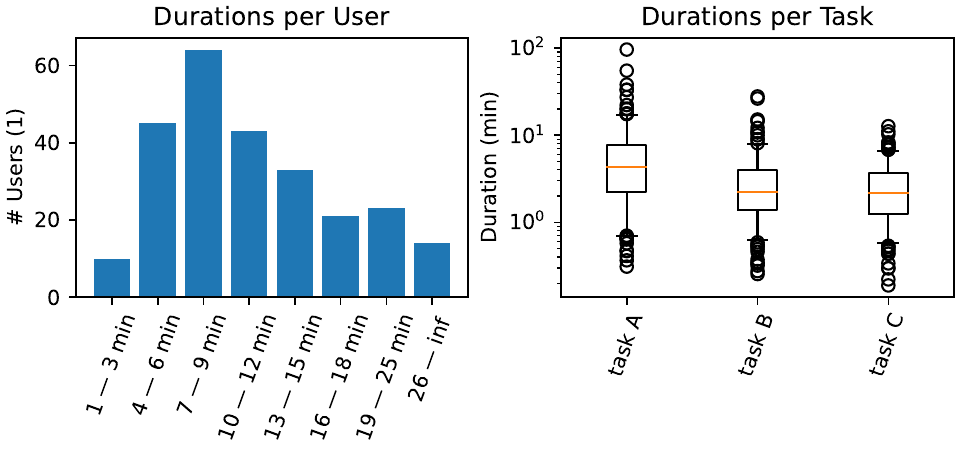} \\
    \includegraphics[width=\linewidth]{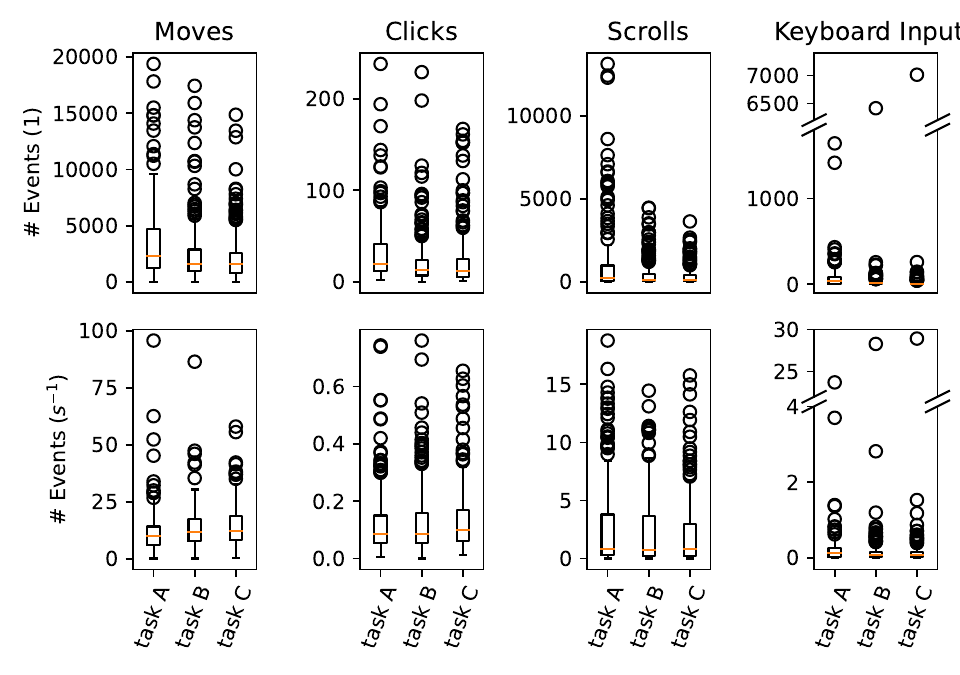}
    \caption{Task-related statistics of overall durations (top) and absolute/relative counts of different events (bottom).}
    \label{fig:task-statistics}
\end{figure}

%-------------------------------------------------------------------------
\subsection{Statistics}\label{sec:statistics}
From 271 users (having completed 798 tasks), 253 (716) remained after the filtering. 
The remaining 716 completed tasks comprise a total of 2,911 separate sessions (see Sec.~\ref{sec:data-structure}), containing a total of $2.39$~M mouse and keyboard events. 
For all users, we also collected sociodemographic information regarding age, gender, education, handedness, and amblyopia, as shown in the break down in Fig.~\ref{fig:sociodemographic-statistics}. 
The majority of participants belong to the youngest age group (18-30 years). 
Note that we required participants to 
(i) be at least 18 years of age; 
(ii) use a desktop setup (i.e., monitor with mouse and keyboard); and 
(iii) not suffer from a visual impairment. 
In terms of gender, we observe a 2:1 majority of female participants and the majority received a secondary or even tertiary education. 
On the conducted \gls{dkt2} test on Diabetes Mellitus on average participants were able to answer $9.42$ out of $14$ questions correctly ($\mathit{SD}=1.95$), exhibiting a Normal distribution.

In terms of the duration required for solving the tasks, we observe a wide spread among subjects and tasks, as outlined in Table~\ref{tab:statistics} and Fig.~\ref{fig:task-statistics}. 
The overall duration for solving all tasks follows a Poisson distribution ($\textrm{mean}=7.79$ minutes, $\textrm{median}=9.90$ minutes), with two outliers (user IDs~104 and 106) with a duration of 55 and 96 minutes for \taskA, respectively. 
An inspection of the respective interaction logs reveals that in both cases there are gaps in the data accounting for almost the whole duration. 
As the exploration view was active during this time, we reason that the respective subjects took a break from the study.

\bgroup
\begin{table}[ht!]
    \centering
    \caption{Fraction of correct answers to the \gls{ir} tasks and the experienced \gls{icl} and \gls{ecl} according to the questions by Krieglstein et al.~\cite{krieglstein2023development}.
    Values in parentheses show the standard deviation. 
    }
    \label{tab:cl}
    \begin{tabular}{l@{\hspace{1pt}}|@{\hspace{1pt}}l l l l l@{}}
        \hline 
            & & \multicolumn{1}{c}{\taskA} & \multicolumn{1}{c}{\taskB} & \multicolumn{1}{c}{\taskC} & \multicolumn{1}{c}{$\Sigma$} \\ 
            \hline\hline
        \multicolumn{2}{l}{Correct} & 0.49 & 0.80 & 0.87 & 0.72 \vspace{2mm} \\ 
        \hline 
        \multirow{6}{*}{\rotatebox{90}{\gls{icl}}} & ICL1 & 2.30 (2.23) & 1.95 (2.03) & 1.64 (1.91) & 1.96 (2.08) \\
        & ICL2 & 2.24 (2.25) & 1.79 (1.96) & 1.45 (1.73) & 1.83 (2.02) \\
        & ICL3 & 3.00 (2.41) & 2.95 (2.24) & 2.55 (2.27) & 2.83 (2.32) \\
        & ICL4 & 3.52 (2.47) & 3.21 (2.41) & 2.86 (2.40) & 3.20 (2.44) \\
        & ICL5 & 2.18 (2.27) & 1.97 (2.06) & 1.46 (1.80) & 1.87 (2.07) \\
        \hhline{~=====}
        & $\Sigma$ & 2.65 (2.39) & 2.37 (2.23) & 1.99 (2.12) & 2.34 (2.26) \vspace{2mm} \\
        \hline 
        \multirow{6}{*}{\rotatebox{90}{\gls{ecl}}} & ECL1 & 3.76 (2.77) & 2.88 (2.52) & 2.57 (2.39) & 3.07 (2.61) \\
        & ECL2 & 3.21 (2.57) & 2.37 (2.32) & 2.14 (2.29) & 2.57 (2.44) \\
        & ECL3 & 3.26 (2.64) & 2.51 (2.38) & 2.27 (2.28) & 2.68 (2.48) \\
        & ECL4 & 4.09 (2.86) & 2.76 (2.63) & 2.49 (2.49) & 3.11 (2.76) \\
        & ECL5 & 3.09 (2.70) & 2.63 (2.51) & 2.39 (2.55) & 2.70 (2.61) \\
         \hhline{~=====}
        & $\Sigma$ & 3.48 (2.74) & 2.63 (2.48) & 2.37 (2.41) & 2.83 (2.59) \\ 
        \hline
    \end{tabular}
\end{table}
\egroup
\begin{figure}[ht!]
    \centering
    \includegraphics[width=\linewidth]{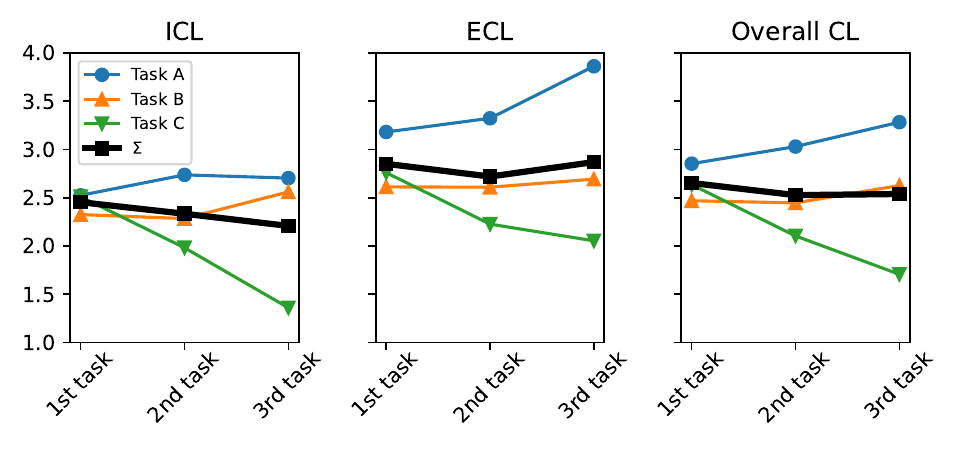}
    \caption{The mean experienced \gls{icl}, \gls{ecl}, and overall \gls{cl} of all subjects over the course of the study. }
    \label{fig:cl-over-time}
\end{figure}

\bgroup
\def\imgWidth{0.24\textwidth}
\def\padding{3pt}
\begin{figure*}[htb]
    \centering
    \subfloat[\taskA]{\includegraphics[width=\imgWidth]{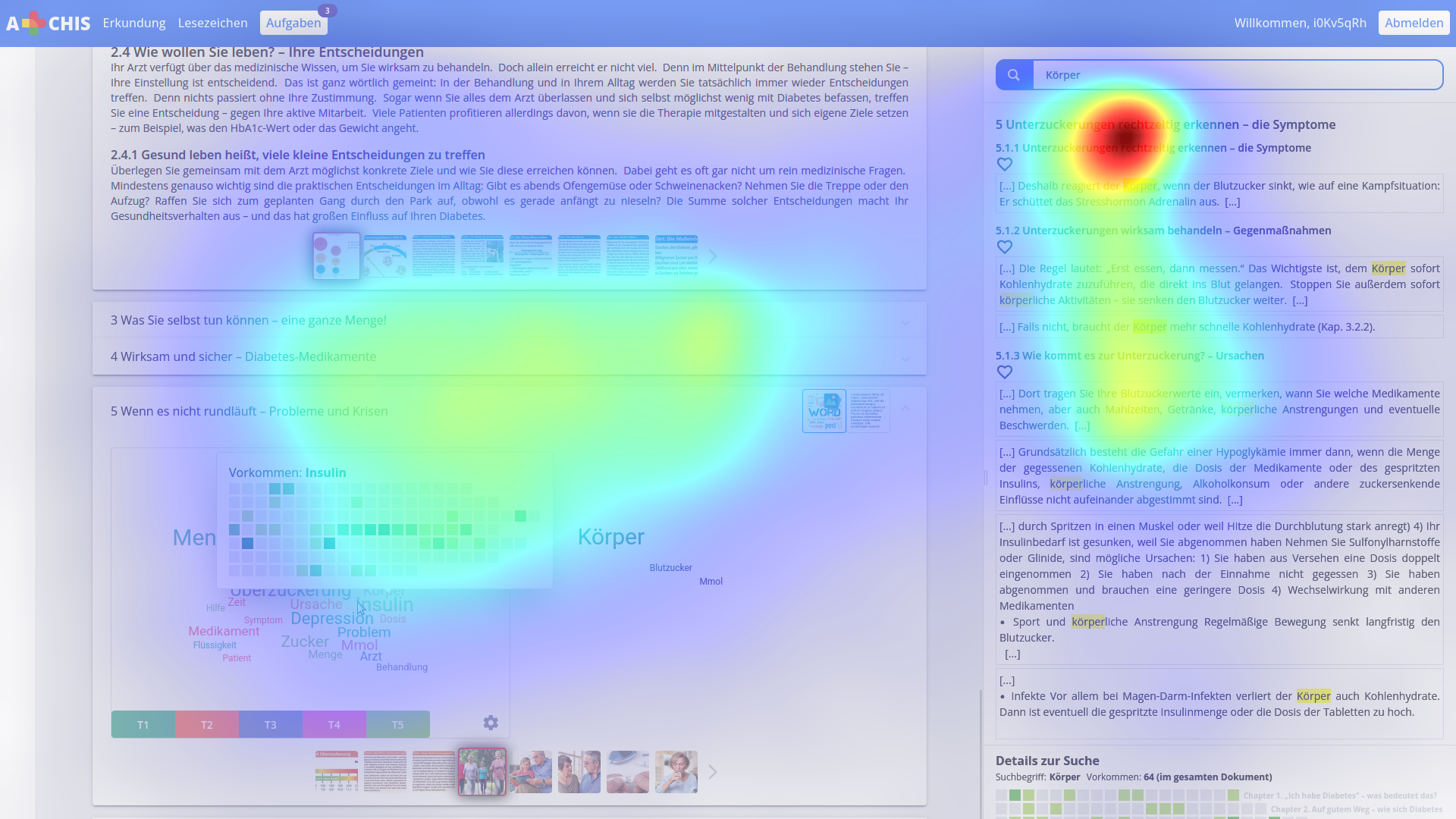}\label{subfig:heatmap:taskA}}\hspace{\padding}
    \subfloat[\taskB]{\label{subfig:heatmap:taskB}\includegraphics[width=\imgWidth]{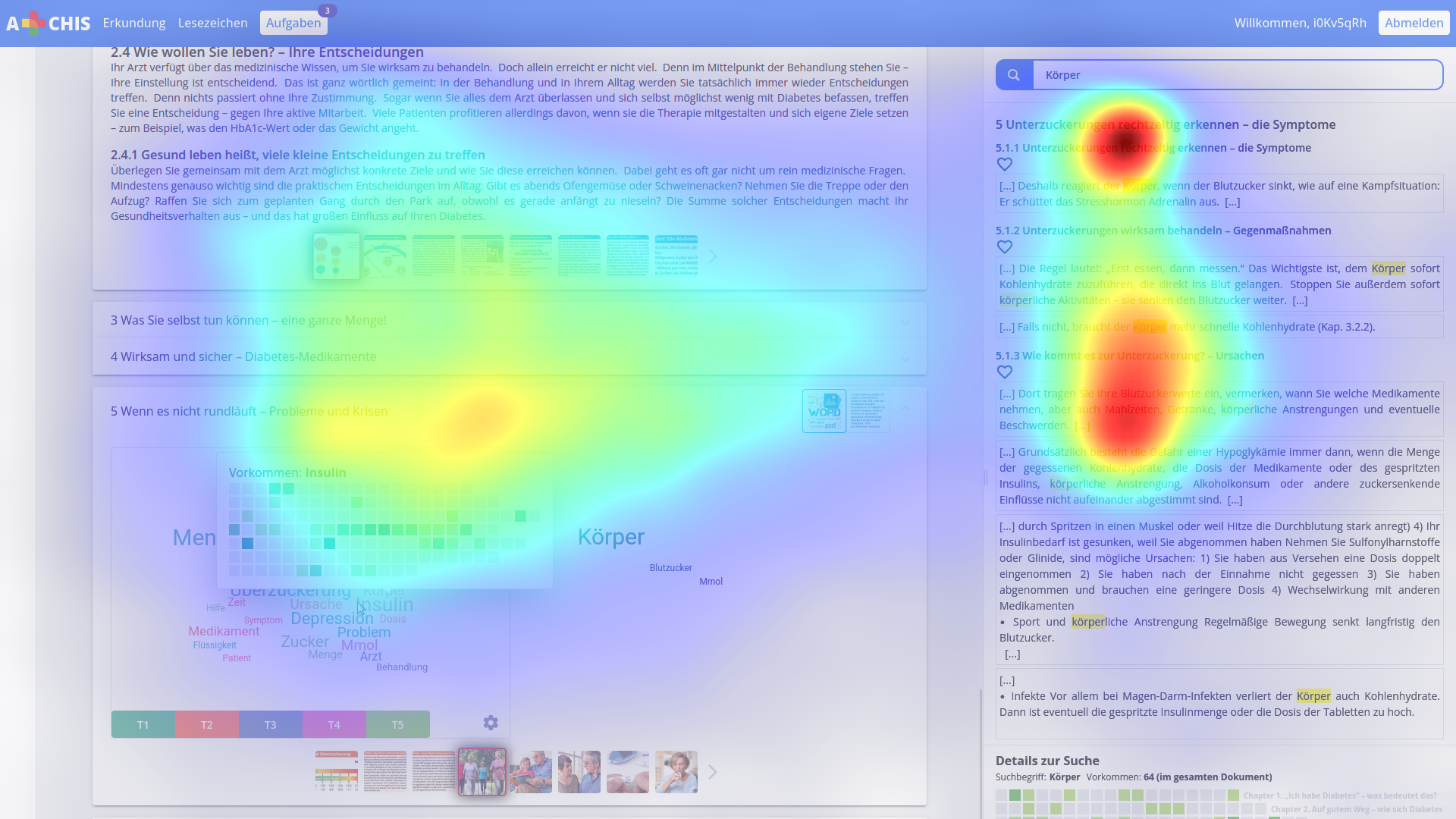}}\hspace{\padding}
    \subfloat[\taskC]{\label{subfig:heatmap:taskC}\includegraphics[width=\imgWidth]{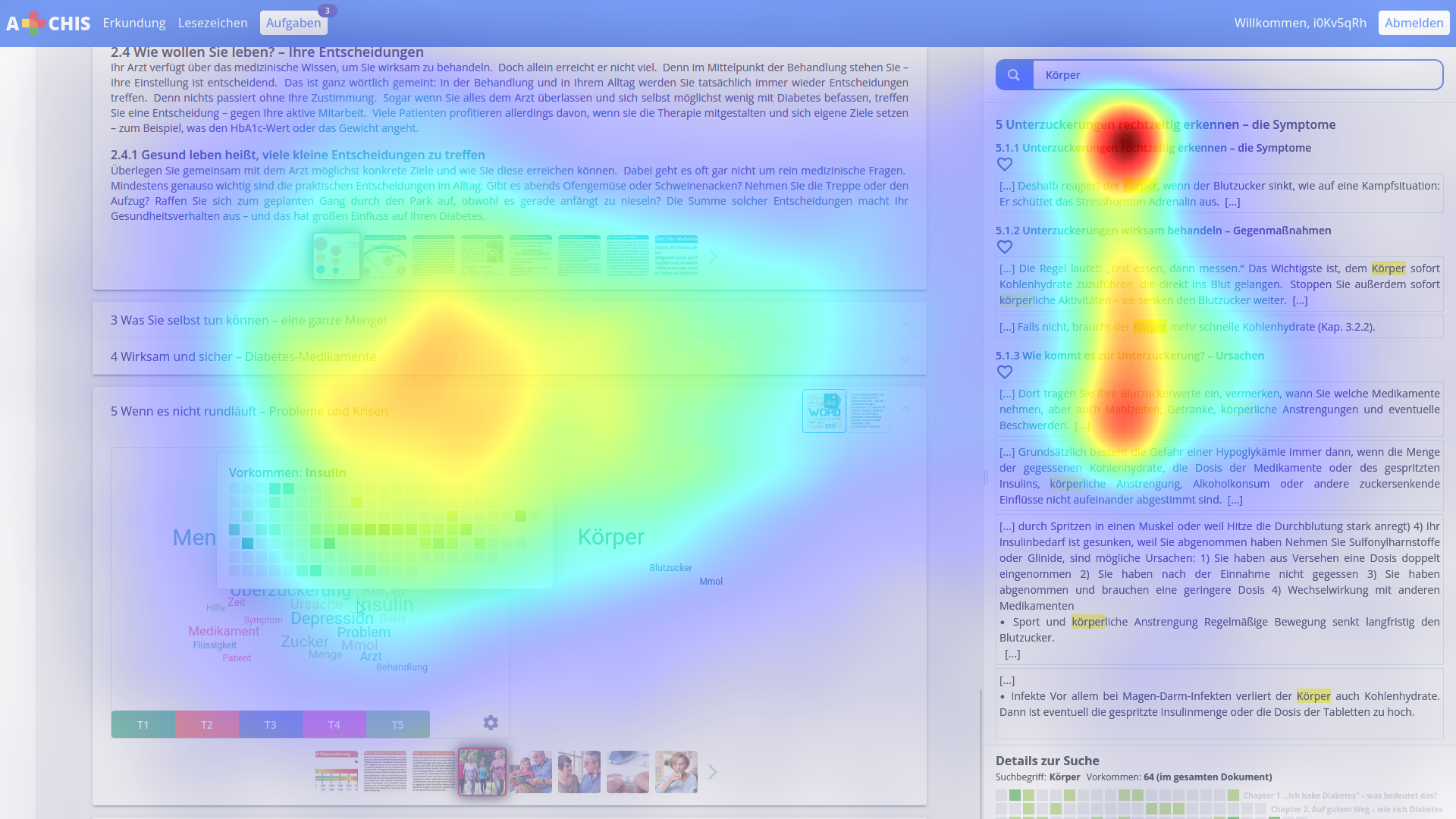}}\hspace{\padding}
    \subfloat[Combined]{\label{subfig:heatmap:combined}\fcolorbox{black}{white}{\includegraphics[width=\imgWidth]{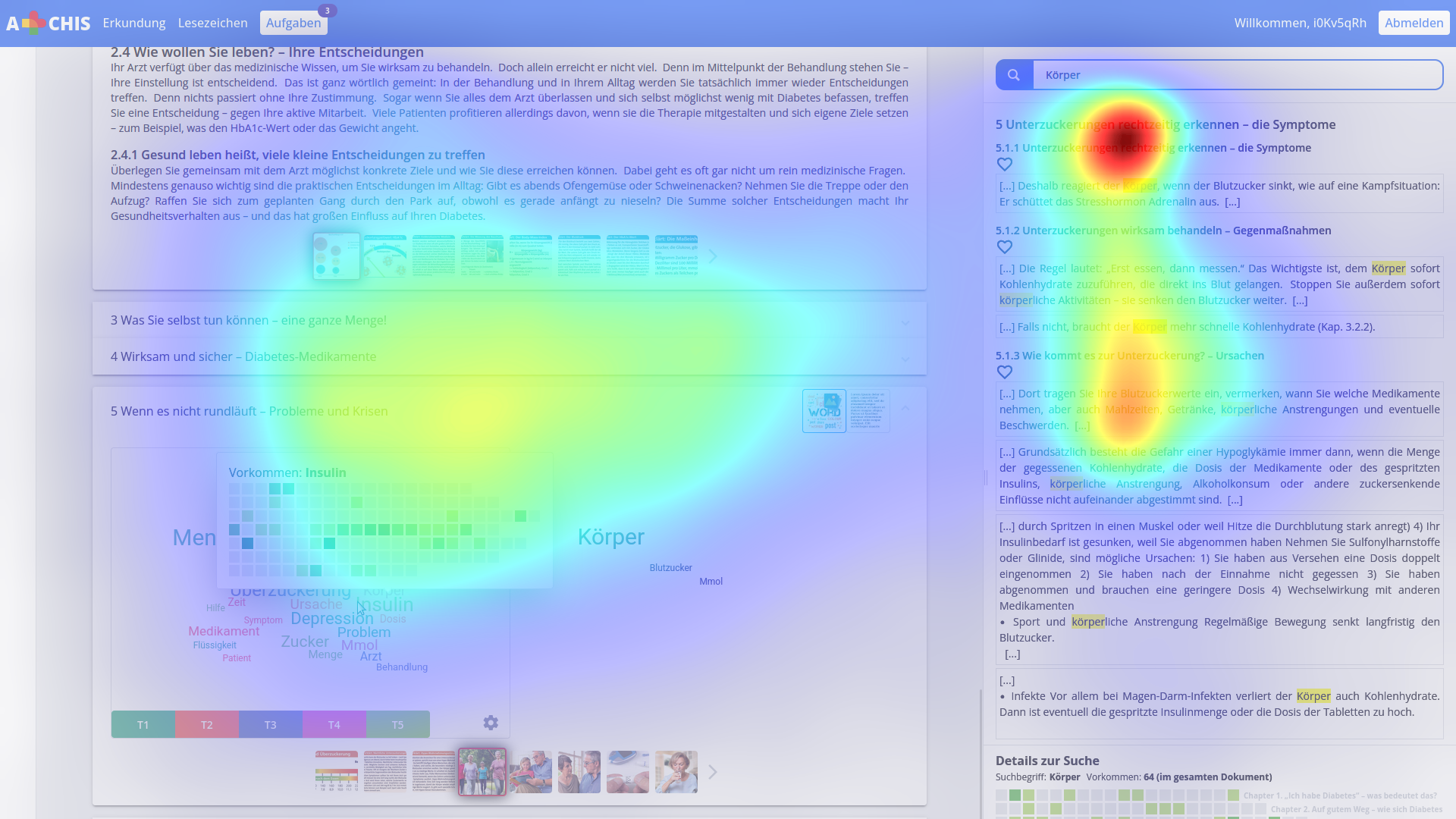}}}
    \caption{Mouse cursor positions for all participants over the course of \taskA--\taskC\ respectively \protect\subref{subfig:heatmap:taskA}--\protect\subref{subfig:heatmap:taskC} and for all tasks combined \protect\subref{subfig:heatmap:combined}. 
    The heatmaps reveal two distinct intensity clusters over the center of the main exploration view and the searchbar with adjacent results list.}
    \label{fig:heatmaps}
\end{figure*}
\egroup

Comparing tasks, we observe that \taskA\ took substantially longer to answer (more than twice the time) than \taskB\ and \taskC. 
It appears that \taskA\ is, unintentionally, much harder to answer than the other two. 
However, in terms of input activity, the tasks are largely balanced as reflected in Fig.~\ref{fig:task-statistics}, bottom. 
Although there appears to be slightly more activity for \taskA\ due to the generally longer durations, this imbalance disappears when looking at the time-normalized statistics. 
In terms of keyboard inputs, there are two users with substantially more events than all others. 
An inspection of the respective logs reveals that one subject (user ID 53) employs the arrow keys for scrolling, and another subject (user ID 153) has continuous and unperturbed input from the 0-key on the numpad, which we attribute to a hardware issue.

The assumption that \taskA\ was harder to solve is also backed-up by the \gls{cl} scores and the correctness rate, reported in Table~\ref{tab:cl}.
We observe that only $49.37\%$ of the participants were able to correctly answer \taskA, while rates were $80.33\%$ and $87.40\%$ for \taskB\ and \taskC, respectively.
The same is true for the \gls{cl} scores. 
Compared to the established questionnaire by Krieglstein et al.~\cite{krieglstein2023development}, we implemented two adjustments: 
First, the subscale on Germane \gls{cl} was left out due to recent theoretical developments of the \gls{cl} theory, which proposes dividing Germane \gls{cl} into \gls{icl} and \gls{ecl}, which in turn add up to a total \gls{cl} score (for a review on recent developments of the \gls{cl} theory see Duran et al.~\cite{duran2022cognitive}). 
Second, we slightly reworded some items. For example, some items include the terms `learning material' or `learning content', which could have been confusing for participants in our context. 
Consequently, we used the terms `material' and `content' (in German: `Material' and `Inhalt'). 
Both the \gls{icl} and \gls{ecl} are substantially higher for \taskA\ and appear to even increase over the course of the study (Fig.~\ref{fig:cl-over-time}).

To get an estimate of the subjects' general cursor patterns, we generate intensity heatmaps for the cursor positions during both the individual tasks (Fig.~\ref{subfig:heatmap:taskA}--\subref{subfig:heatmap:taskC}) and overall (Fig.~\ref{subfig:heatmap:combined}). 
In order to combine mouse trajectories, we normalized all positions (c.f. Sec.~\ref{sec:data-handling}) and scaled them to a common 16:9 aspect ratio and overlayed it to a characteristic view of the system. 
Note that the actually used aspect ratio depends on the subjects' window setup and the appearance of the exploration view is dynamic as sections can be expanded/collapsed and representations changed. 
It is, nonetheless, a good approximation for the majority of participants. 
That is, for all tasks we can observe two intensity clusters reflecting the system's bipartite layout: (1) a bulgy cluster at the center of the scrollable document view, which constitutes the ``main'' component, and (2) a strong i-shaped manifestation on the search interface on the right-hand side. 
The latter can be easily explained through the characteristics of a search process -- i.e., a user rests the mouse over the input field while typing and subsequently moves the cursor downwards over the top results. 
One particularity we can observe is that there is more intense usage of the document view for \taskC\ which we attribute to that fact that the answer for this task can be found solely in the pictorial content, which is displayed there.

% schema created with https://dbdiagram.io/d

% // Use DBML to define your database structure
% // Docs: https://dbml.dbdiagram.io/docs
% Table user {
%   id integer [primary key]
%   created_at timestamp
%   age integer
%   gender enum
%   education enum
%   handedness enum
%   amblyopia enum
%   "[dkt2answers]" char
%   "[dkt2aswer_durations]" integer
%   "[sus_scales]" integer
%   activity float
% }

% Table mouse_event {
%   id integer [primary key]
%   session_id integer [not null]
%   timestamp timestamp
%   type enum
%   client_x integer
%   client_y integer
%   chapter integer
%   component enumerate
%   context json
% }

% Table keyboard_event {
%   id integer [primary key]
%   session_id integer [not null]
%   timestamp timestamp
%   key char
%   modifiers integer
%   context json
% }

% Table task {
%   id integer [primary key]
%   task_number integer [not null]
%   user_id integer [not null]
%   submit_ts timestamp
%   correctness float
%   answer string
%   "[cognitive_load]" integer
% }

% Table session {
%   id integer [primary key]
%   task_id integer [not null]
%   from_ts timestamp
%   to_ts timestamp
%   initialization_type enum
%   finalziation_type enum
%   window_size json
% }

% Table image {
%   id integer [primary key]
%   chapter integer
%   page_number integer
%   caption string
% }

% Table sentence {
%   id integer [primary key]
%   chapter string
%   content string
% }

% Ref user: task.user_id > user.id 
% Ref task: session.task_id > task.id 
% Ref session: mouse_event.session_id > session.id
% Ref session: keyboard_event.session_id > session.id

\begin{figure*}[htpb!]
    \centering
    \includegraphics[width=\textwidth]{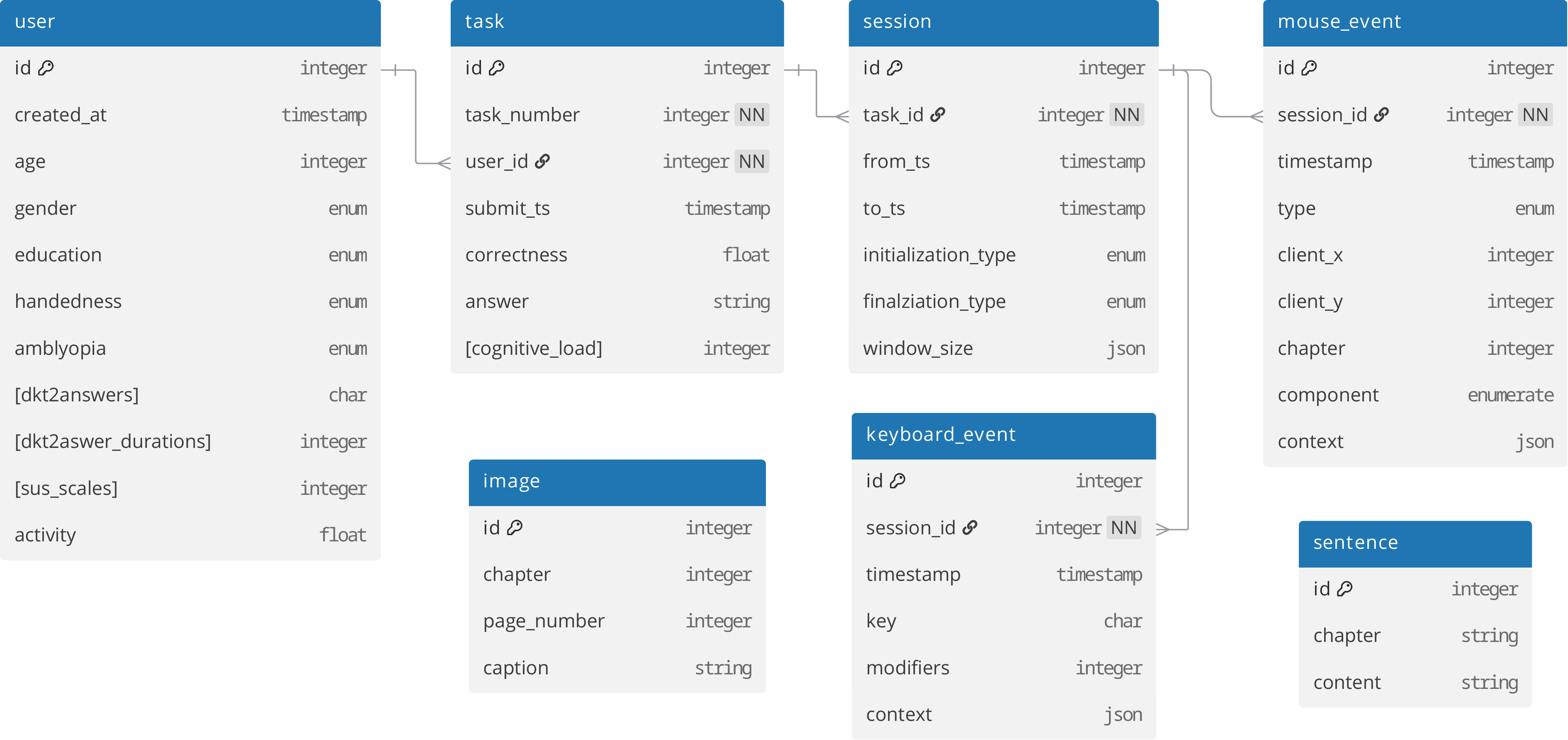}
    \caption{The underlying schema of the \gls{guideta}.
    Note, that \texttt{image} and \texttt{sentence} tables are connected to the \texttt{keyboard\_event}/\texttt{mouse\_event} tables via their \texttt{context} attribute in which the respective IDs appear.
    }
    \label{fig:schema}
\end{figure*}

%-------------------------------------------------------------------------
\subsection{Data Structure and Organization}\label{sec:data-structure}

The presented dataset was drawn from the \gls{apchis}'s database, removing all bloat which is not directly related to the behavioral data or relevant for further research. 
Further steps were taken to comply with the \gls{fair} Guiding Principles for scientific data management and stewardship~\cite{wilkinson2016fair}. 
Those include a normalization of the used identifiers. 
As many of the raw data records were not part of the study or subjected to filtering, the thereof resulting fragmented ID ranges were pruned accordingly. 
All timestamps, which were mostly recorded in GMT+2 time zone, were transformed to UTF and converted to floating-point Unix timestamps. 
We organize the data in tabular scheme (Fig.~\ref{fig:schema}), adhering to the second normal form of relational databases~\cite{codd1972further}. 

We differentiate between two types of inputs (henceforth referred to as \textbf{InputEvents}), triggered by the user:
\begin{description}
    \item [KeyboardEvent] is recorded upon observing a ``keydown''\footnote{\url{https://developer.mozilla.org/en-US/docs/Web/API/Element/keydown_event}} event. 
    Additional to the time of the event, we log the key which has been pressed and the modifiers which have been active during the keypress. 
    The latter allows us to infer the usage of shortcuts. 
    If a type-in element (e.g., a search bar) was in focus at the time of the event, we also log the current input string of this element.     
    \item [MouseEvent] describes a mouse click, -move, or -scroll event. 
    Besides a timestamp, it comprises the cursor's X/Y coordinates in viewport space\footnote{\url{https://developer.mozilla.org/en-US/docs/Web/CSS/CSSOM_view/Coordinate_systems}}, the entire viewing area in which the document is presented in the browser. 
    Note that the value ranges for these coordinates are different from user to user as they depend on the browsers' window sizes on their respective monitors. 
    Even for a certain user they can change at any moment if the window is resized or altered in another way. 
    Consequently, we also store the viewport sizes at all times (see \textbf{Session}) and also provide software for normalizing the mouse trajectories (Sec.~\ref{sec:data-handling}). 
\end{description}

\begin{figure}[ht!]
    \centering
    \includegraphics[]{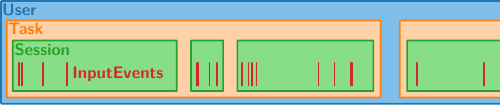}
    \caption{All \textbf{InputEvents} are contained within three hierarchical containers, referring to a specific \textbf{User}, \textbf{Task}, and time frame of uninterrupted exploration (i.e., \textbf{Session}).}
    \label{fig:dataframe-relations}
\end{figure}

All \textbf{InputEvents} are organized in three tiers of containers, as illustrated in Fig.~\ref{fig:dataframe-relations}, which allow to relate them to a specific \textbf{User}, \textbf{Task}, and \textbf{Session}:
\begin{description}
    \item [User] is the uppermost container and comprises all user-related information, such as sociodemographic parameters, \gls{sus} answers, \gls{dkt2} answers, or parameters describing the overall activity. 
    \item [Task] is the data frame for the posed \gls{ir} tasks (up to three per user) and features the given answer, whether it was correct, and the \gls{cl} indicated by the user through a filled-out questionnaire. 
    \item [Session] is the lowest container and comprises the actual \textbf{InputEvents}. 
    This additional tier is necessary since the IR task can be interrupted -- either by switching to another (browser) window or other predefined actions which indicate an interruption of the exploration process, i.e., opening the help wizard or task modal. 
    Besides the start and end timestamp of a \textbf{Session}, we store its initialization- and finalization context, e.g., ``window loosing focus'' or ``help wizard closed''. 
    As the study was conducted purely unsupervised these session time frames together with their initialization/finalization contexts provide vital cues for detecting illicit behavior, such as using other sources for answering the posed tasks or revealing whether users experience issues with the platform's components (indicated through frequent help wizard visits). 
    That is, this information can be employed for filtering based on additional inclusion criteria. 
    A \textbf{Session} also contains the window sizes as key value-pairs with timestamp, necessary for normalizing the cursor positions. 
    Usually the window size remains constant over the course of a \textbf{Session}, but could change if a user, e.g., changes to full screen mode.
\end{description}

The schema in Fig.~\ref{fig:schema} also features a table \textbf{Sentence}, which contains an index list of sentences extracted from the data source~\cite{baumgartDiabetesImGriff2021}, which we use in \gls{apchis}. 
\textbf{Sentences} can be linked to some of the \textbf{InputEvents} if they contain a reference in their respective contexts (Sec.~\ref{sec:event-contexts}). 
Detailed descriptions of the additional attributes of the schema are documented in the software framework (Sec.~\ref{sec:data-handling}), which we provide for further processing.
All data records are publicly available on OSF (\url{https://osf.io/fhvbm/}).
To ensure interoperability~\cite{wilkinson2016fair} without any custom software, we provide them as CSV files, with the following folder structure, relating to the naming conventions in Fig.~\ref{fig:schema}:
\dirtree{%
.1 /.
.2 mouse\_events.
.3 me\_<user>\_<task>.csv.
.2 keyboard\_events.
.3 ke\_<user>\_<task>.csv.
.2 task\_answers.csv.
.2 sessions.csv.
.2 users.csv.
.2 sentences.csv.
.2 images.csv.
}

%-------------------------------------------------------------------------
\subsubsection{Event Contexts}\label{sec:event-contexts}

\begin{figure}[ht!]
    \centering
    \includegraphics[width=\linewidth]{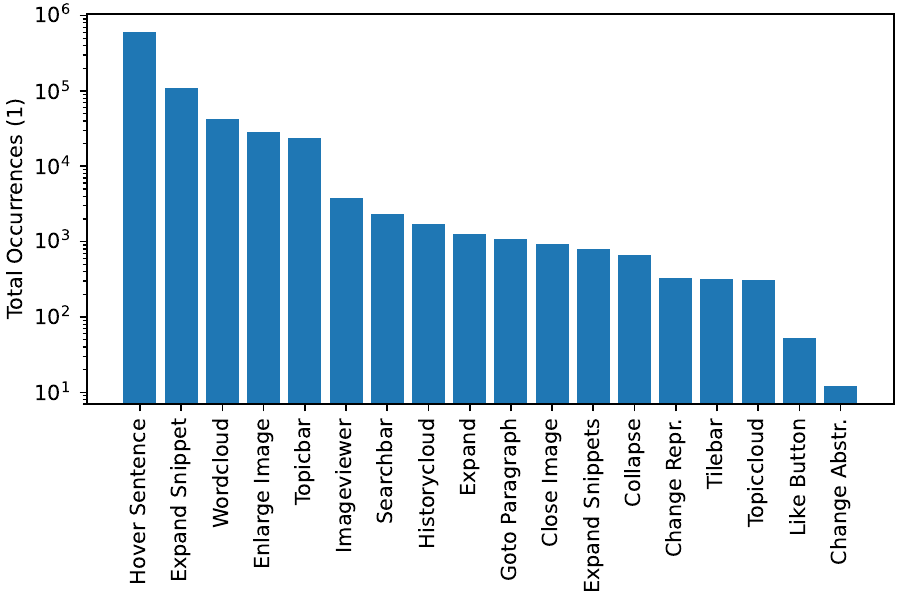}
    \caption{The 18 predefined actions supported by our system, together with their respective prevalence in the \gls{guideta} dataset.}
    \label{fig:contexts-distribution}
\end{figure}

Several \textbf{MouseEvents} ($827$~K, i.e., $35.13\%$) feature additional context information. 
Besides widget information (the reference to the currently used visual component) and content information (the reference to the currently displayed information unit), we also log specific types of interaction with the system. 
To this end, we predefined a set of 18 actions which are supported by our system. 
On a high-level they include things like enlarging content, different interactions with the system's visual components, navigation, bookmarking, and searching. 
Details on the different actions, which also store related context information (e.g., the particular word, which has been clicked in case of a click in a word cloud), are documented within our data handling software framework (Sec.~\ref{sec:data-handling}).
Fig.~\ref{fig:contexts-distribution} shows a breakdown of the occurrences of the different actions.
The, by far, most prevalent action is \emph{Hover Sentence} -- the event that a user hovers or clicks a particular sentence -- followed by expanding text snippets, clicking or hovering a term in the word cloud, enlarging an image, etc. 
These context data allow to, e.g., 
(i) determine if users perform clicks, which are not supported by the system; 
(ii) identify reoccurring interaction cycles; or 
(iii) determine a preference for certain tools. 

\begin{figure*}[ht!]
    \centering
    \includegraphics[width=\linewidth]{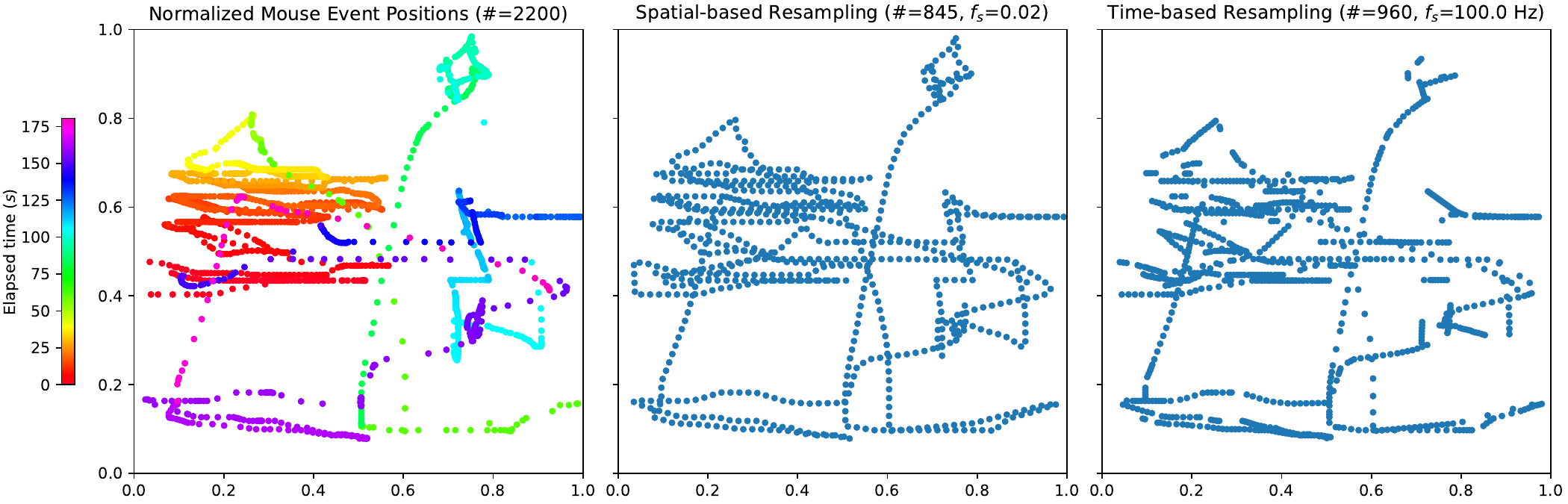}
    \caption{
    Left: Normalized mouse positions for a randomly selected exploration session of approximately 3 minutes. 
    Middle: The spatial-based resampling (with a sampling frequency $f_s$ of 2\% of the screen space's diagonal) is able to faithfully capture the overall trajectory while reducing the number of positions by more than half. 
    Right: With time-based resampling ($f_s = 100~\mathrm{Hz}$), some trajectory features could get lost, while it is still beneficial for some applications.}
    \label{fig:resample-example}
\end{figure*}

%-------------------------------------------------------------------------
\section{Data Handling and Analysis}\label{sec:data-handling}

Alongside the described dataset, we also publish the software framework\footnote{\url{https://github.com/lenxn/apchis-guidaeta.git}} used for processing the data. 
Written in Python, it allows to easily 
(i) load all the data records in Fig.~\ref{fig:schema} as class objects, 
(ii) filter for specific records (e.g., `get all \textbf{MouseEvents} associated with \taskC'), and 
(iii) normalize positions.
The latter is relevant for \textbf{MouseEvents} whose coordinates are stored in viewport space. 
Together with the \texttt{window\_size} attribute in their belonging \textbf{Session} they can be mapped to a $[0,1)$ value domain (for X and Y independently, or in combined fashion).

Based on the \texttt{context} and widget (i.e., \texttt{component}) information in the \textbf{InputEvents}, we define two higher-level interaction types. 
First, a \textbf{Dwelling}, which constitutes an unperturbed interaction with a certain visual component for a period of time. 
As such, it has start- and end timestamp as well as information pertaining to the used widget and the displayed content. 
Second, a \textbf{Hovering}, which is an interval of an unperturbed hovering over a certain element. 
This differs from a \textbf{Dwelling} in the sense that it is a transient interaction which can be interrupted by clicks, scrolls, or keyboard inputs. 
\textbf{Hoverings} are thus often enclosed within a \textbf{Dwelling}. 
We note that these interactions merely serve as a starting point for behavior modeling or other purposes and can be adjusted or extended depending on given requirements.

In an effort to capture as much information as possible, our collected mouse positions stem from an event-driven sampling scheme -- i.e., the HTML events triggered by the used browsers. 
This results in a very high sampling rate (up to 95 Hz, c.f. Table~\ref{tab:statistics}) in most cases.
However, events are unevenly spaced in both time and space. 
For various trajectory analyses, the (re)sampling of the cursor positions constitutes a necessary prerequisite. 
In general, we differentiate between spatial-based or time-based normalization~\cite{kieslich_mouse-tracking_2019}. 
With the prior, the mouse trajectory is represented by a series of locations with uniform distances. 
Likewise, with the latter, the mouse trajectory is represented by a series of locations evenly spaced in time. 
The time-based normalization is beneficial for the comparison of short-time actions~\cite{spivey_continuous_2005} while the spatial-based normalization is advantageous for the spatial analysis of taken paths and long-term processes. 
Both approaches are implemented in our software framework (Fig.~\ref{fig:resample-example}), in which the user is able to provide custom sample frequencies or distances. 
The framework was also employed to generate all statistics figures shown in this paper.

%-------------------------------------------------------------------------
\section{Potential Limitations and Future Work}\label{sec:discussion}
We believe that the published dataset constitutes a valuable contribution for interaction-driven studies in different application domains. 
Nonetheless, we want to point out potential issues which potentially limit the usage of the data for specific purposes. 
Even though we strive for diversity in terms of gender, background, and ages in our set of participants, the subjects which actually took part in the study are predominantly from a young age group, female and of high education backgrounds (Fig.~\ref{fig:sociodemographic-statistics}). 
This has to be considered when interpreting behavioral patterns and drawing conclusions regarding generalizability.

The unsupervised nature of the study setup enabled a broad outreach and allowed us to efficiently collect data from hundreds of participants. 
On the downside, this left us with limited control over the hardware and environment setup subjects used to conduct the study. 
While some properties, such as the window sizes over the duration of the study are closely monitored and logged, other factors, such as the mouse sensitivity, double-click threshold, dpi, and all properties pertaining to the look and feel of the system are beyond our control. 
That means that -- although we log which components and contents were displayed at a specific time -- we cannot faithfully reconstruct the actual representation seen by the user.

The context information (Sec.~\ref{sec:event-contexts}), associated to \textbf{InputEvents}, is strongly tailored to the \gls{apchis}, with clearly defined event types such as `expanding a chapter' or `bookmarking a certain piece of information'. 
While this limits the generalizability of our interaction tracking (which was purposely developed for this study), it provides very detailed information on how an exploration process was conducted, painting a very clear picture of users' behavior. 
We can, e.g., evaluate the time a user spends on processing a patch of information as opposed to the time spent on finding new information patches -- something that is studied within the \emph{information foraging theory}~\cite{pirolliInformationForaging1999}.

Regarding future work, we also intend to use the dataset to study the effectiveness of different behavioral indicators for measuring cognitive load. 
Also, building upon previous work~\cite{acm-umap-2024}, we develop and evaluate visual interfaces, which enable expert users to effectively review behavioral patterns and reveal between-subject similarities~\cite{xuSurveyAnalysisUser2020b}. 
Beyond that, the dataset is well-suited for experimenting with user classification and development of \gls{cl} over time. 
As we also report the correctness of the given task answers (reflecting users' success), it even allows to study which factors/interactions lead to either a successful completion or a failure.

%-------------------------------------------------------------------------
\section{Conclusion}
With the \gls{guideta}, we present an encompassing dataset obtained from 253 individual users. 
The core strengths -- besides the sheer volume -- are the rich context and content information associated with most \textbf{InputEvents}, as well as various additional metadata pertaining to sociodemographic information about participants, answers from various feedback questionnaires, and experienced \gls{cl}. 
We believe that, together with the tools for data handling and processing, \gls{guideta} can be employed in versatile fashion and constitutes thus a valuable contribution to different research domains.

%-------------------------------------------------------------------------
\section*{Acknowledgments}
This work was funded by the Austrian Science Fund (FWF) as part of the project `Human-Centered Interactive Adaptive Visual Approaches in High-Quality Health Information' (\gls{apchis}; Grant No. FG 11-B).

%-------------------------------------------------------------------------
\bibliographystyle{eg-alpha-doi} 
\bibliography{egbibsample}  

\end{document}